\documentclass[11pt]{article}
\usepackage{indentfirst}
\usepackage{epsfig}
\usepackage{amssymb,amsmath,amsfonts,mathrsfs}
\usepackage{subfigure}
\usepackage{cite}
\usepackage{float}

\newenvironment{mysummary}[1]{%
    \leftskip=2.5em \rightskip=2.5em
    \noindent\small{\bfseries #1}}
    {\par\medskip}
\renewenvironment{abstract}{\begin{mysummary}{Abstract:}}{\end{mysummary}}
\newenvironment{keywords}{\begin{mysummary}{Key words:}}{\end{mysummary}\medskip}
\newenvironment{proof}{\par{\itshape Proof}.\ }
    {\hfill\raisebox{.56ex}{\fbox{}}\par}

\newtheorem{theorem}{\indent Theorem}

\newtheorem{myremark}{\indent Remark}
\newenvironment{remark}{\begin{myremark}\normalfont}
    {\end{myremark}}
\newtheorem{myexample}{\indent Example}

\title{Three-Dimensional Integrated Guidance and Control Based on Small-Gain Theorem\thanks{This work was supported by National Natural Science Foundation of China (No.61203125) and Fundamental Research Funds for the Central Universities (No. HIT. NSRIF. 2013039).}}
\author{Han Yan\footnote{Engineer, e-mail address: yhbice@gmail.com.}\\[1ex]
Science and Technology on Space
Intelligent Control Laboratory,\\
Beijing Institute of Control Engineering,
Beijing 100190, China\\[1ex] and\\ Mingzhe Hou\footnote{Lecturer, e-mail address: hithyt@gmail.com.}\\[1ex]Center for Control Theory and Guidance Technology,\\ Harbin Institute of Technology, Harbin 150001, China}
\date{}

\begin{document}

\maketitle

\begin{abstract}
A three-dimensional (3D) integrated guidance and control (IGC) design approach is proposed by using small-gain theorem in this paper. The 3D IGC model is formulated by combining nonlinear pursuer dynamics with the nonlinear dynamics describing pursuit-evasion motion. Small-gain theorem and ISS theory are iteratively utilized to design desired attack angle, sideslip angle and attitude angular rates (virtual controls), and eventually an IGC law is proposed. Theoretical analysis shows that the IGC approach can make the LOS rate converge into a small neighborhood of zero, and the stability of the overall system can be guaranteed as well.
\end{abstract}
\begin{keywords}
Three-dimensional integrated guidance and control; Generalized small-gain theorem; Input-to-state stability; Robustness.
\end{keywords}

\section*{Nomenclature}
\noindent\begin{tabular}{@{}lcl@{}}
\textit{$\alpha$} &=& angle of attack \\
\textit{$\beta$} &=& angle of sideslip \\
\textit{$\gamma$} &=& roll angle \\
\textit{$\vartheta$} &=& pitch angle \\
\textit{$\omega_{i}~(i=x,y,z)$} &=& body-axis roll, yaw and pitch rates \\
\textit{$\delta_{i}~(i=x,y,z)$} &=& aileron, rudder and elevator deflections \\
\textit{$V$} &=& velocity of the pursuer \\
\textit{$m$} &=& mass of the pursuer \\
\textit{$P$}  &=&  thrust force\\
\textit{$\rho$} &=& air density\\
\textit{$q =0.5qV^2$} &=& dynamic pressure\\
\textit{$J_{i}~(i=x,y,z)$} &=& roll, yaw and pitch moments of inertia\\
\textit{$X,Y,Z$} &=& drag, lift and side forces\\
\textit{$S,L$} &=& reference area, reference length\\
\textit{$r$} &=& relative range between pursuer and evader\\
\textit{$\theta_{L},\varphi_{L}$} &=& LOS elevation, LOS azimuth\\
\textit{$\theta_{V},\varphi_{V}$} &=& velocity elevation, velocity azimuth\\
\textit{$c_{x0}$} &=& zero-lift drag coefficient\\
\textit{$F_{i}~(i=V,\theta,\varphi)$} &=& force components along the axes of the velocity coordinate system\\
\textit{$c_{x}^{\alpha},c_{x}^{\beta}$} &=& partial derivatives of drag force coefficient with respect to $\alpha$ and $\beta$\\
\textit{$c_{x}^{\delta_{x}},c_{x}^{\delta_{y}},c_{x}^{\delta_{z}}$} &=& partial derivatives of drag force coefficient with respect to $\delta_{x}$, $\delta_{y}$ and $\delta_{z}$\\
\textit{$c_{x}^{\alpha\beta}$} &=& second partial derivatives of drag force coefficient with respect to $\alpha$ and $\beta$\\
\textit{$c_{y}^{\alpha}$,$c_{y}^{\beta}$,$c_{y}^{\delta_{z}}$} &=& partial derivatives of lift force coefficient with
respect to $\alpha$, $\beta$ and $\delta_{z}$\\
\textit{$c_{z}^{\alpha}$,$c_{z}^{\beta}$,$c_{y}^{\delta_{y}}$} &=& partial derivatives of side force coefficient with
respect to $\alpha$, $\beta$ and $\delta_{y}$\\
\textit{$m_{x}^{\delta_{x}}$,$m_{x}^{\alpha}$,$m_{x}^{\beta}$} &=& partial derivatives of rolling moment coefficient
with respect to $\delta_x$, $\alpha$ and $\beta$\\
\textit{$m_{y}^{\beta}$,$m_{y}^{\delta_{y}}$} &=& partial derivatives of yawing moment coefficient
with respect to $\beta$ and $\delta_{y}$\\
\textit{$m_{z}^{\alpha}$,$m_{z}^{\delta_{z}}$} &=& partial derivatives of pitching moment coefficient
with respect to $\alpha$ and $\delta_{z}$
\end{tabular} \\

\section{Introduction}
The guidance and control systems of vehicles are usually designed separately, and in order to achieve the desired overall system performance, modifications are generally inevitably required to each subsystem. Hence, the traditional design approach usually leads to excessive design iterations and high costs. What's more, strictly speaking, the stability of the overall system cannot be guaranteed \cite{Adaptive_block_dynamic_surface_control}. Integrated guidance and control (IGC) design is regarded as one of emerging trends in vehicle control technology, because it views guidance and control loops as an integrated system and taking couplings between subsystems into account, and besides that, such a design can reduce the cost of the required sensors and increase the system reliability \cite{inegrated_guidance_control_first}. Due to those reasons, IGC design has received more and more attention recently.

After IGC design was put forward in \cite{inegrated_guidance_control_first}, various control methods have been introduced, and sliding-mode control (SMC) is a typical method, which is used in most of the existing relevant literatures to solve the two-dimensional IGC design problem for the pursuit-evasion game. The second-order SMC was used to design IGC laws in \cite{SOSM_guidance_control_2009} and \cite{finite_time_inegrated_guidance_control_JGCD}. In \cite{SOSM_guidance_control_2009}, a sliding surface that depends on the line-of-sight (LOS) rate was defined in the guidance loop with the pursuer pitch rate viewed as a virtual control, and the second-order SMC was used to control the pitch rate to track the virtual control robustly in finite time. For pursuers steered by a combination of aerodynamic lift, sustainer thrust, and center-of-gravity divert thrusters, an IGC algorithm, integrated with the smooth second-order sliding mode guidance law in \cite{SSOSM-based_guidance_law}, was developed using second-order SMC to achieve an accurate tracking of the attitude command \cite{finite_time_inegrated_guidance_control_JGCD}. Similarly, \cite{WXH_2013} also designed the pitch rate command in outer loop, and the inner loop was constructed to track the outer loop command, where the finite time convergence can be guaranteed in both two loops according to the novel adaptive nonsingular terminal SMC method proposed in the paper. Shima and co-workers used SMC to obtain IGC approaches for pursuers with only one control input \cite{sliding-mode_gl_2006} and pursuers with both canard and tail controls \cite{sliding-mode_gl_2007} with the assumption that the evader acceleration can be measured.
Note that, in order to remove nonlinear terms, the equations of IGC model in \cite{sliding-mode_gl_2006} and
\cite{sliding-mode_gl_2007} were all formulated under the assumption that the angle between LOS and pursuer velocity is almost constant, but it might be not proper in practice since large maneuvers of a evader may lead to significant variation of that angle. The IGC laws in \cite{Yan_IGC_backstepping} and \cite{small_gain_theorem_for_IGC} were proposed without that assumption. To deal with the nonlinear terms, an adaptive control method was introduced into the backstepping scheme to design an IGC law \cite{Yan_IGC_backstepping}. For dual-control pursuers, small-gain theorem \cite{Small-GainTheorem} was also used to design IGC law in \cite{small_gain_theorem_for_IGC} to enforce the attitude angle (rate) commands that are aimed at producing desired aerodynamic lift to achieve robust tracking of a maneuvering evader. Both IGC laws in \cite{Yan_IGC_backstepping} and \cite{small_gain_theorem_for_IGC} can make the LOS rate converge into a small neighborhood of zero in the presence of evader maneuvers and pursuer model uncertainties.

Actually, an actual pursuit-evasion motion occurs in a three-dimensional (3D) environment. Only when the couplings between lateral and normal motion are ignored, the design and analysis of IGC laws can be simplified into two planar relative motions. However, such an approach is ad hoc in nature, and the 3D IGC law design is a challenging problem.

For IGC problem in three dimensions, some nonlinear optimal control methods, such as state dependent Riccati equation (SDRE) technique \cite{Riccati_for_inegrated_guidance_control_by_ Menon,Riccati_for_inegrated_guidance_control_by_ palumbo} and $\theta-D$ technique \cite{theta-D_method_for_inegrated_guidance_control}, were utilized. These methods all involve complicated numerical computations since the Hamilton-Jaccobi-Bellman (HJB) equation is needed to be solved on-line and that is time consuming. What is more, these methods cannot ensure the robustness of the closed-loop system. Without complicated numerical computations, adaptive block dynamic surface control, which can avoid ``explosion of complexity'' problem when comparing with backstepping method, was used to design 3D IGC law \cite{Adaptive_block_dynamic_surface_control}. A set of first-order filters were introduced at each step of the traditional block backstepping approach, and the stability analysis of the closed-loop system was also given based on the Lyapunov theory. But similarly to \cite{sliding-mode_gl_2006} and \cite{sliding-mode_gl_2007}, \cite{Adaptive_block_dynamic_surface_control} also assumed that the angle between LOS and pursuer velocity is constant.

All the works mentioned above made great contributions to the development of IGC design, but many existing results were obtained based on some strong assumptions or without considering robustness against uncertainties and disturbances. In addition, most of 3D IGC laws involve complicated numerical computations and cannot stabilize the overall system.

In this paper, a novel 3D IGC design approach is proposed for skid-to-turn (STT) vehicles by iteratively using small-gain theorem and input-to-state stability (ISS) \cite{Smooth_stabilization_implies_coprime_factorization}. The desired attack angle and sideslip angle are designed to make the LOS rate be ISS with respect to evader maneuvers. Then, by iteratively utilizing small-gain theorem, the desired attitude angular rates and the final IGC law are proposed to drive the attack angle and sideslip angle to track their commands. Theoretical analysis show that the IGC approach makes both the LOS rate and the tracking error of attitude angle (rate) be input-to-state practically stable (ISpS) with respect to evader maneuvers and pursuer model uncertainties. It is worth to claim that our approach is formulated considering the couplings between lateral and pitch channels, and the nonlinearity caused by the moving between LOS and pursuer velocity is also taken into consideration. Besides, the stability of the overall system can be guaranteed by small-gain theorem, and comparing with the backstepping scheme, the procedures of our design approach do not involve the derivatives of virtual controls, such that the problem of ``explosion of complexity'' is avoided.

The remainder of this paper is organized as follows. The 3D integrated guidance and control model is formulated in Section \ref{sec of model}. After presenting some basic concepts, the IGC law is designed in Section \ref{sec of the IGC law design}, and also stability of the overall pursuer system is analyzed. Finally, Section \ref{Conclusions} summarizes the conclusions.
\section{Model Derivation}\label{sec of model}
The nonlinear pursuer dynamics with uncertainties proposed in \cite{Adaptive_block_dynamic_surface_control} is described by
\begin{subequations}\label{original_model}
\begin{equation}
\dot{x}_{1}=f_{1}(x_{1})+g_{1}(\vartheta,x_{1})x_{2}+d_{1}
\end{equation}
\begin{equation}
\dot{x}_{2}=f_{2}(x_{1},x_{2})+g_{2}(t)u+d_{2}
\end{equation}
\end{subequations}
where
\begin{subequations}
\begin{equation}\nonumber
x_{1}=\begin{bmatrix}
\gamma\\
\alpha\\
\beta
\end{bmatrix},~x_{2}=\begin{bmatrix}
\omega_{x}\\
\omega_{y}\\
\omega_{z}
\end{bmatrix},~u=\begin{bmatrix}
\delta_{x}\\
\delta_{y}\\
\delta_{z}
\end{bmatrix},
\end{equation}
\begin{equation}\nonumber
f_{1}(x_{1})=\begin{bmatrix}
0\\
-\frac{1}{mV\cos\beta}\left(P\sin\alpha+qSC_{y}^{\alpha}\alpha\right)\\
\frac{1}{mV}\left(qSC_{z}^{\beta}\beta-P\cos\alpha\sin\beta\right)
\end{bmatrix},~g_{1}(\vartheta,x_{1})=\begin{bmatrix}
1 & -\tan\vartheta\cos\gamma & \tan\vartheta\sin\gamma\\
-\tan\beta\cos\alpha & \sin\alpha\tan\beta & 1\\
\sin\alpha & \cos\alpha & 0
\end{bmatrix},
\end{equation}
\begin{equation}\nonumber
f_{2}(x_{1},x_{2})=\begin{bmatrix}
\frac{J_{z}-J_{y}}{J_{x}}\omega_{y}\omega_{z}\\
\frac{1}{J_{y}}qSLm_{y}^{\beta}\beta+\frac{J_{x}-J_{z}}{J_{y}}\omega_{x}\omega_{z}\\
\frac{1}{J_{z}}qSLm_{z}^{\alpha}\alpha+\frac{J_{y}-J_{x}}{J_{z}}\omega_{x}\omega_{y}
\end{bmatrix},~g_{2}(t)=\begin{bmatrix}
\frac{1}{J_{x}}qSLm_{x}^{\delta_{x}} & 0 & 0\\
0 & \frac{1}{J_{y}}qSLm_{y}^{\delta_{y}} & 0\\
0 & 0 & \frac{1}{J_{z}}qSLm_{z}^{\delta_{z}}
\end{bmatrix}
\end{equation}
\end{subequations}
and
\begin{equation}
\dot{\vartheta}=\omega_{y}\sin\gamma+\omega_{z}\cos\gamma
\end{equation}
where $d_{1}$ and $d_{2}$ are uncertainties.

\begin{figure}[H]\centering
\begin{minipage}{12cm}
\centering
\epsfig{file=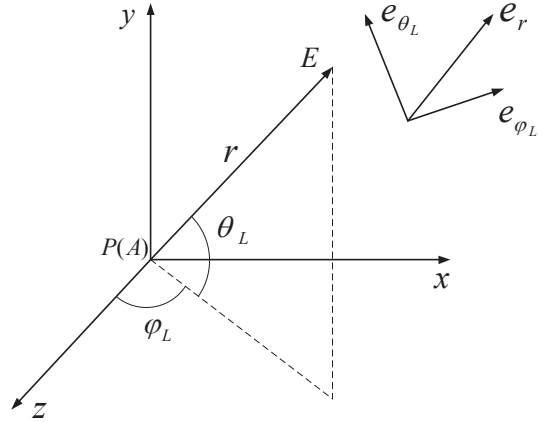,height=6cm} \caption{The Pursuit-Evasion Motion\label{The Pursuit-Evasion Motion}}
\end{minipage}
\end{figure}

Consider the spherical LOS coordinates $(r,\theta_{L},\varphi_{L})$ with origin fixed
at the pursuer's gravity center. As shown in Fig. \ref{The Pursuit-Evasion Motion}, let $(\mathbf{e_{r}},\mathbf{e_{\theta_{L}}},\mathbf{e_{\varphi_{L}}})$ be the unit vectors
along the coordinate axes, $r$ be the relative range between pursuer and evader, $\theta_{L}$ be the LOS elevation, and $\varphi_{L}$ be the LOS azimuth. The components of the relative acceleration is given as \cite{Adaptive_block_dynamic_surface_control,Yan_HGOGL}

\begin{subequations}\label{exmodel}
\begin{equation}\label{equation of r}
\ddot{r}=r(\dot\varphi_{L})^{2}\cos^{2}\theta_{L}+r(\dot\theta_{L})^{2}+a_{E_{r}}-a_{P_{r}}
\end{equation}
\begin{equation}\label{equation of theta}
\ddot{\theta}_{L}=\frac{-2\dot{r}\dot{\theta}_{L}-r(\dot{\varphi}_{L})^{2}\cos\theta_{L}\sin\theta_{L}+a_{E_{\theta_{L}}}-a_{P_{\theta_{L}}}}{r}
\end{equation}
\begin{equation}\label{equation of phi}
\ddot{\varphi}_{L}=\frac{-2\dot{r}\dot{\varphi}_{L}}{r}+2\dot{\varphi}_{L}\dot{\theta}_{L}\tan\theta_{L}+\frac{a_{E_{\varphi_{L}}}-a_{P_{\varphi_{L}}}}{r\cos\theta_{L}}
\end{equation}
\end{subequations}
where $(a_{P_{r}},a_{P_{\theta_{L}}},a_{P_{\varphi_{L}}})$ and $(a_{E_{r}},a_{E_{\theta_{L}}},a_{E_{\varphi_{L}}})$ are, respectively, the acceleration vectors of pursuer and evader in
the LOS coordinate system.

\begin{figure}[H]\centering
\begin{minipage}{12cm}
\centering
\epsfig{file=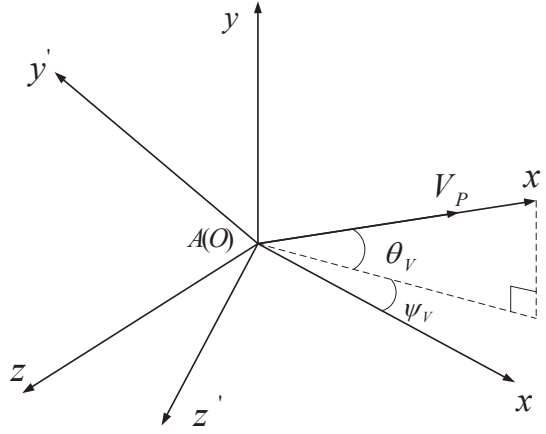,height=6cm} \caption{Ground Coordinate System $(Axyz)$ and Pursuer Velocity Coordinate System $(Ox^{'}y^{'}z^{'})$\label{dd_dm}}
\end{minipage}
\end{figure}

The relationship between ground coordinate system and pursuer velocity coordinate system is shown in Fig. \ref{dd_dm}, where $Ox^{'}$ axe is along the pursuer velocity vector, $\theta_{V}$ is the velocity elevation, and $\psi_{V}$ is the velocity azimuth. Let $(F_{V},F_{\theta},F_{\psi})$ be the force components along the axes of the velocity coordinate system, and one has \cite{book_by_Li}
\begin{subequations}\label{pursuer point motion}
\begin{equation}
ma_{V}=m\frac{\mathrm{d}V}{\mathrm{d}t}=F_{V}
\end{equation}
\begin{equation}
ma_{\theta}=mV\frac{\mathrm{d}\theta_{V}}{\mathrm{d}t}=F_{\theta}
\end{equation}
\begin{equation}
ma_{\psi}=-mV\cos\theta_{V}\frac{\mathrm{d}\psi_{V}}{\mathrm{d}t}=F_{\psi}
\end{equation}
\end{subequations}
where $(a_{V},a_{\theta},a_{\psi})$ is the acceleration vector of the pursuer in velocity coordinate system.
The vector $(x,y,z)$ in the ground coordinate system can be transformed to the pursuer velocity coordinate system through the following equation \cite{book_by_Li}
\begin{equation}\label{transform between Axyz and Ox2y2z2}
\begin{bmatrix}
x'\\
y'\\
z'
\end{bmatrix}=\mathbf{L}(\psi_{V},\theta_{V})
\begin{bmatrix}
x\\
y\\
z
\end{bmatrix}
\end{equation}
where
\begin{equation}\nonumber
\mathbf{L}(\psi_{V},\theta_{V})=
\begin{bmatrix}
\cos\theta_{V}\cos\psi_{V} & \sin\theta_{V} & -\cos\theta_{V}\sin\psi_{V}\\
-\sin\theta_{V}\cos\psi_{V} & \cos\theta_{V} & \sin\theta_{V}\sin\psi_{V}\\
\sin\psi_{V} & 0 & \cos\psi_{V}
\end{bmatrix}
\end{equation}

Therefore, according to Eq. (\ref{transform between Axyz and Ox2y2z2}) and the definitions of the velocity elevation and azimuth, one can obtain
\begin{equation}\label{transform for acc}
\begin{bmatrix}
a_{P_{r}}\\
a_{P_{\theta_{L}}}\\
-a_{P_{\varphi_{L}}}
\end{bmatrix}=\mathbf{L}\left(\varphi_{L}-\frac{\pi}{2},\theta_{L}\right)\mathbf{L}^{-1}(\psi_{V},\theta_{V})
\begin{bmatrix}
a_{V}\\
a_{\theta}\\
a_{\psi}
\end{bmatrix}
\end{equation}
In practical applications, during the end game, the pursuer speed is usually assumed to be constant, i.e., $a_{V}=0$ \cite{small_gain_theorem_for_IGC,sliding-mode_gl_2006,sliding-mode_gl_2007}.

The acceleration components of the pursuer along the y- and z-axes of the pursuer velocity coordinate system are given by \cite{book_by_Li}
\begin{equation}
\begin{bmatrix}
a_{\theta}\\
a_{\psi}
\end{bmatrix}=\frac{1}{m}
\begin{bmatrix}
P\sin\alpha+Y\\
-P\cos\alpha\sin\beta+Z
\end{bmatrix}
\end{equation}
where lift force $Y$ and side force $Z$ are given by
\begin{subequations}
\begin{equation}
Y=qSC_{y}^{\alpha}\alpha+d_{y}
\end{equation}
\begin{equation}
Z=qSC_{z}^{\beta}\beta+d_{z}
\end{equation}
\end{subequations}
with uncertainties $d_{y}$ and $d_{z}$. When $\alpha$ and $\beta$ are small enough, we have $\sin\alpha\approx\alpha$, $\sin\beta\approx\beta$ and $\cos\alpha\approx1$. Thus,
\begin{equation}\label{a_theta(alpha and beta)}
\begin{bmatrix}
a_{\theta}\\
a_{\psi}
\end{bmatrix}=\frac{1}{m}
\begin{bmatrix}
P+qSC_{y}^{\alpha} & 0\\
0 & -P+qSC_{z}^{\beta}
\end{bmatrix}\begin{bmatrix}
\alpha\\
\beta
\end{bmatrix}+\underbrace{\frac{1}{m}\begin{bmatrix}
d_y\\
d_z
\end{bmatrix}}_{d_{V}}
\end{equation}
From Eq. (\ref{transform for acc}), we have
{\small\begin{equation}\label{transform for a_M and a_theta}
\begin{bmatrix}
a_{P_{\theta_{L}}}\\
a_{P_{\varphi_{L}}}
\end{bmatrix}=\begin{bmatrix}
\sin\theta_{L}\sin\theta_{V}\sin(\varphi_{L}-\psi_{V})
+\cos\theta_{L}\cos\theta_{V} & -\sin\theta_{L}\cos(\psi_{V}-\varphi_{L})\\
-\sin\theta_{V}\cos(\varphi_{L}-\psi_{V}) & -\sin(\varphi_{L}-\psi_{V})
\end{bmatrix}
\begin{bmatrix}
a_{\theta}\\
a_{\psi}
\end{bmatrix}\triangleq M(t)\begin{bmatrix}
a_{\theta}\\
a_{\psi}
\end{bmatrix}
\end{equation}}
Define
\begin{equation}\label{definition of variables}
x_{01}=\omega_{\theta}=\dot{\theta}_{L},~x_{02}=\omega_{\phi}=\dot{\phi}_{L}\triangleq\dot{\varphi}_{L}\cos\theta_{L}
\end{equation}
and
\begin{equation}
x_{0}=\begin{bmatrix}
x_{01}\\
x_{02}
\end{bmatrix},~x_{1}^{\#}=\begin{bmatrix}
\alpha\\
\beta
\end{bmatrix}
\end{equation}
From (\ref{exmodel}), (\ref{a_theta(alpha and beta)}) and (\ref{transform for a_M and a_theta}), we have
\begin{equation}
\dot{x}_{0}=f_{0}(x_{0})+g_{0}(t)x_{1}^{\#}+\frac{d_{0}(t)}{r}
\end{equation}
where
\begin{equation}
f_{0}(x_{0})=\begin{bmatrix}
-2\frac{V_{r}}{r}x_{01}-x_{02}^{2}\tan\theta_{L}\\
-2\frac{V_{r}}{r}x_{02}+x_{01}x_{02}\tan\theta_{L}
\end{bmatrix},~g_{0}(t)=-\frac{M(t)}{mr}\begin{bmatrix}
P+qSC_{y}^{\alpha} & 0\\
0 & -P+qSC_{z}^{\beta}
\end{bmatrix}
\end{equation}
and $d_{0}(t)=-M(t)d_{V}+\begin{bmatrix}a_{E_{\theta_{L}}}
\\a_{E_{\varphi_{L}}}\end{bmatrix}$ is assumed to be bounded disturbance.

According to the above analysis, the IGC model can
be written as
\begin{subequations}\label{IGC model}
\begin{equation}\label{IGC model 1}
\dot{x}_{0}=f_{0}(x_{0})+g_{0}(t)x_{1}^{\#}+\frac{d_{0}}{r}
\end{equation}
\begin{equation}\label{IGC model 2}
\dot{x}_{1}=f_{1}(x_{1})+g_{1}(\vartheta,x_{1})x_{2}+d_{1}
\end{equation}
\begin{equation}\label{IGC model 3}
\dot{x}_{2}=f_{2}(x_{1},x_{2})+g_{2}(t)u+d_{2}
\end{equation}
\end{subequations}

Let $(\mathbf{i}_{x},\mathbf{i}_{y},\mathbf{i}_{z})$ be the unite vectors along the ground coordinate axes, and we can see from Fig. (\ref{dd_dm}) that
\begin{subequations}
\begin{equation}
\textbf{\textsl{r}}=r\cos\theta_{L}\sin\varphi_{L}\textbf{\textsl{i}}_{x}+r\sin\theta_{L}\textbf{\textsl{i}}_{y}+r\cos\theta_{L}\cos\varphi_{L}\textbf{\textsl{i}}_{z} \end{equation}
\begin{equation}
\textbf{\textsl{V}}_{M}=V_{M}\cos\theta_{V}\cos\psi_{V}\textbf{\textsl{i}}_{x}+V_{M}\sin\theta_{V}\textbf{\textsl{i}}_{y}-V_{M}\cos\theta_{V}\sin\psi_{V}\textbf{\textsl{i}}_{z}
\end{equation}
\end{subequations}
It is easy to verify that
\begin{equation}
\det(M(t))=\frac{\textbf{\textsl{r}}\cdot\textbf{\textsl{V}}_{M}}{r^{2}V_{M}}
\end{equation}
holds, so when pursuer velocity is orthogonal onto the LOS, we have $\det(M(t))=0$, that is, $M(t)$ is non-invertible in this case. But the angle between LOS and pursuer velocity is always acute in the whole process of homing guidance \cite{Yan_IGC_backstepping}, thus we assume that the matrix $M(t)$ is invertible here, and in this case, $g_{0}(t)$ is invertible. Due to the analysis of \cite{Adaptive_block_dynamic_surface_control}, if $\alpha$, $\beta$ and $\vartheta$ are all kept in a reasonable domain around zero, $g_{1}(\vartheta,x_{1})$ is also invertible for arbitrary variable $\gamma$, so we assume that $g_{1}(\vartheta,x_{1})$ is invertible in a reasonable flight domain.

\section{Integrated Guidance and Control Law Design}\label{sec of the IGC law design}
In this section, small-gain theorem and ISS theory are iteratively used to design desired attack angle, sideslip angle and attitude angular rates (virtual controls), and eventually an IGC law is proposed. Theoretical analysis shows that the IGC approach can make the LOS rate converge into a small neighborhood of zero, and the stability of the overall system can be guaranteed as well.

For any measurable function $u(t):\mathbb{R}_{+}\rightarrow \mathbb{R}^{m}$, $\|u(t)\|_{s}$ denotes $\sup\limits_{0\leq \tau\leq t}\|u(\tau)\|$.
\subsection{Concepts and Preliminaries}
Consider the following general interconnected system
\begin{equation}\label{interconnected system1}
H_{1}:~\dot{x}_{1}=f_{1}(x_{1},y_{2},u_{1}),~y_{1}=h_{1}(x_{1},y_{2},u_{1})
\end{equation}
\begin{equation}\label{interconnected system2}
H_{2}:~\dot{x}_{2}=f_{2}(x_{2},y_{1},u_{2}),~y_{2}=h_{2}(x_{2},y_{1},u_{2})
\end{equation}
where, for $i=1,2$, $x_{i}\in \mathbb{R}^{n_{i}}$, $u_{i}\in \mathbb{R}^{m_{i}}$, and $y_{i}\in \mathbb{R}^{p_{i}}$. The functions $f_{1},f_{2},h_{1}$ and $h_{2}$ are smooth and a smooth function $h$ exists such that
\begin{equation}\nonumber
(y_{1},y_{2})=h(x_{1},x_{2},u_{1},u_{2}),
\end{equation}
is the unique solution of
\begin{eqnarray}\nonumber
\left\{
\begin{aligned}
y_{1}=h_{1}(x_{1},h_{2}(x_{2},y_{1},u_{2}),u_{1})\\
y_{2}=h_{2}(x_{2},h_{1}(x_{1},y_{2},u_{1}),u_{2})
\end{aligned}\right.
\end{eqnarray}
We have:
\begin{theorem}\label{Small-Gain Theorem}
\cite{Small-GainTheorem}
Suppose (\ref{interconnected system1}) and (\ref{interconnected system2}) are input-to-state stability (ISS) with $(y_{2},u_{1})$ $($respectively $(y_{1},u_{2})$$)$ as input, $y_{1}$ $($respectively $y_{2}$$)$ as output, and there exist class $\mathcal {K}\mathcal {L}$ functions $\beta_{1}$, $\beta_{2}$, class $\mathcal {K}$ functions $\gamma_{1y}$, $\gamma_{1u}$, $\gamma_{2y}$, $\gamma_{2u}$, and nonnegative constants $d_{1}$, $d_{2}$ such that
\begin{eqnarray}\nonumber
\left\{
\begin{aligned}
\|y_{1}(t)\|\leq \beta_{1}(\|x_{1}(0)\|,t)+\gamma_{1y}(\|y_{2}(t)\|_{s})+\gamma_{1u}(\|u_{1}(t)\|_{s})+d_{1}\\
\|y_{2}(t)\|\leq \beta_{2}(\|x_{2}(0)\|,t)+\gamma_{2y}(\|y_{1}(t)\|_{s})+\gamma_{2u}(\|u_{2}(t)\|_{s})+d_{2}
\end{aligned}\right.
\end{eqnarray}
If two class $\mathcal{K_{\infty}}$ functions $\rho_{1}$ and $\rho_{2}$ and a nonnegative real number $s_{l}$ satisfying
\begin{eqnarray}\label{requirement of small-gain theorem}
\left\{
\begin{aligned}
(Id+\rho_{2})\circ\gamma_{2y}\circ(Id+\rho_{1})\circ\gamma_{1y}(s)\leq s\\
(Id+\rho_{1})\circ\gamma_{1y}\circ(Id+\rho_{2})\circ\gamma_{2y}(s)\leq s
\end{aligned}\right., ~~\forall s\geq s_{l}
\end{eqnarray}
exist, system (\ref{interconnected system1})-(\ref{interconnected system2}) with $u=(u_{1},u_{2})$ as input, $y=(y_{1},y_{2})$ as output and $x=(x_{1},x_{2})$ as state will be input-to-output practically stability (IOpS) $($input-to-output stability (IOS) if $s_{l}=d_{1}=d_{2}=0$$)$.
\end{theorem}

\subsection{ISS-Based Control Law Design}
Consider general nonlinear system
\begin{equation}\label{general nonlinear system}
\dot{x}=f(x,t)+g(x,t)u+d(t)
\end{equation}
where $f:[0,\infty)\times R^{n}\rightarrow R^{n}$, $f:[0,\infty)\times R^{n}\rightarrow R^{n\times n}$ and disturbance $d:[0,\infty)\rightarrow R^{n}$. The following theorem holds.
\begin{theorem}\label{theorem for ISS control law}
Assume $g(x,t)$ is invertible. The closed-loop system of system (\ref{general nonlinear system}) and control law
\begin{equation}\label{ISS-based control law}
u=g^{-1}\left(-f-kx-\frac{1}{2\delta^{2}}x\right)
\end{equation}
is ISS with respect to $d$ for $k>0$ and $\delta>0$, that is,
\begin{equation}\label{inequality in ISS form for theorem 2}
\|x(t)\|\leq e^{-kt}\|x(0)\|+\frac{\delta}{\sqrt{2k}}\sqrt{1-e^{-2kt}}\|d(t)\|_{s}
\end{equation}
Moreover, if disturbance $d$ vanishes, the origin of the closed-loop system will be exponentially stable.
\end{theorem}
\begin{proof}
The derivative of $V=\frac{1}{2}x^{T}x$ along the trajectories of system (\ref{general nonlinear system}) is given by
\begin{equation}\label{dot(V) for general nonlinear system}
\dot{V}=x^{T}(f(x,t)+g(x,t)u+d(t))
\end{equation}
Applying
\begin{equation}
x^{T}d\leq \frac{1}{2\delta^{2}}\|x\|^{2}+\frac{\delta^{2}}{2}\|d\|^{2}
\end{equation}
where $\delta>0$, into (\ref{dot(V) for general nonlinear system}), we obtain
\begin{equation}\label{dot(V) for general nonlinear system with inequality}
\dot{V}\leq x^{T}\left(f(x,t)+g(x,t)u+\frac{1}{2\delta^{2}}x\right)+\frac{\delta^{2}}{2}\|d\|^{2}
\end{equation}
Substituting (\ref{ISS-based control law}) into Eq. (\ref{dot(V) for general nonlinear system with inequality}) yields
\begin{equation}
\dot{V}\leq -k\|x\|^{2}+\frac{\delta^{2}}{2}\|d\|^{2}
\end{equation}
Solving the differential inequality yields
\begin{equation}
V(x(t))\leq e^{-2kt}V(x(0))+\frac{\delta^{2}}{4k}(1-e^{-2kt})\|d(t)\|_{s}^{2}
\end{equation}
Taking the square roots and using the inequality $\sqrt{a^{2}+b^{2}}\leq a+b$ for nonnegative numbers $a$ and $b$, we can see that Eq. (\ref{inequality in ISS form for theorem 2}) holds. Therefore, the closed-loop system of system (\ref{general nonlinear system}) and control law (\ref{ISS-based control law}) is ISS with respect to disturbance $d$.

Moreover, if disturbance $d$ vanishes, that is,
$d=0$, Eq. (\ref{inequality in ISS form for theorem 2}) can be rewritten as $\|x(t)\|\leq e^{-kt}\|x(0)\|$. In
this case, the origin of the closed-loop system is exponentially stable.
\end{proof}

Theorem \ref{theorem for ISS control law} shows that, with the control law (\ref{ISS-based control law}), $x$ can converge to a small neighborhood of zero by adjusting coefficients $k$ and $\delta$ for bounded disturbance $d$.
\subsection{IGC Law Design}
Consider subsystem (\ref{IGC model 1}). $\frac{d_{0}}{r}$ is not bounded when $r=0$, however, due to the finite size of pursuers and evaders, a successful interception can be achieved as long as $r$ decreases to a particular intercept value in the whole process of homing guidance. Thus, $\frac{d_{0}}{r}$ is bounded and we assume that inequality $0<r_{m}<r<r_{M}$ holds \cite{SSOSM-based_guidance_law,Yan_HGOGL}. Since the assumption that $g_{0}$ is invertible is reasonable as analyzed, according to Theorem \ref{theorem for ISS control law}, taking the virtual control law\footnote{The terms $-x_{02}^{2}\tan\theta_{L}$ and $x_{01}x_{02}\tan\theta_{L}$ in $f_{0}$ (the cross couplings between the elevation and the azimuth of LOS) need no consideration when designing the virtual control, see \cite{Yan_HGOGL}.}
\begin{equation}\label{virtual control law 1}
x_{1}^{\#}=g_{0}^{-1}\left(2\frac{V_{r}}{r}-\frac{1}{2\delta^{2}_{0}}-K_{0}\right)x_{0}\triangleq x_{1}^{\#*}
\end{equation}
with $K_{0}>0$ and $\delta_{0}>0$, we can obtain
\begin{equation}\label{inequality of LOS rate in the ISS form}
\|x_{0}(t)\|\leq e^{-K_{0}t}\|x_{0}(0)\|+\frac{\delta_{0}}{\sqrt{2K_{0}}r_{m}}\sqrt{1-e^{-2K_{0}t}}\|d_{0}(t)\|_{s}
\end{equation}
For STT vehicles, the roll angle should be kept near zero throughout the engagement, thus, let $x_{1}^{*}=[0,(x_{1}^{\#*})^{T}]^{T}$, and the change of variables
\begin{equation}
\eta_{1}^{\#}=x_{1}^{\#}-x_{1}^{\#*},\eta_{1}=x_{1}-x_{1}^{*}
\end{equation}
brings Eqs. (\ref{IGC model 1})-(\ref{IGC model 2}) into the form
\begin{eqnarray}\nonumber
H_{1}:\left\{
\begin{aligned}
\dot{x}_{0}&=f_{0}+g_{0}x_{1}^{\#*}+\frac{d_{0}}{r}+y_{1}\\
y_{0}&=-\dot{x}_{1}^{*}
\end{aligned}\right.
\end{eqnarray}
and
\begin{eqnarray}\nonumber
H_{2}:\left\{
\begin{aligned}
\dot{\eta}_{1}&=f_{1}+g_{1}x_{2}+d_{1}+y_{0}\\
y_{1}&=g_{0}\eta_{1}^{\#}
\end{aligned}\right.
\end{eqnarray}
With $x_{1}^{\#*}$, system $H_{1}$ is ISS with respect to $d_{0}$ and $y_{1}$, and due to Eq. (\ref{inequality of LOS rate in the ISS form}), we have
\begin{alignat}{1}
\|x_{0}(t)\|\leq  \underbrace{e^{-K_{0}t}\|x_{0}(0)\|}_{\beta^{x}_{0}(\|x_{0}(0)\|,t)}+\underbrace{\frac{\delta_{0}}{\sqrt{2K_{0}}r_{m}}\sqrt{1-e^{-2K_{0}t}}\|d_{0}(t)\|_{s}}_{\alpha_{0}^{x}(\|d_{0}(t)\|_{s})}+\underbrace{\frac{\delta_{0}}{\sqrt{2K_{0}}}\sqrt{1-e^{-2K_{0}t}}\|y_{1}(t)\|_{s}}_{r_{m}\alpha_{0}^{x}(\|y_{1}(t)\|_{s})}\label{inequality of LOS rate in the ISS form in H1}
\end{alignat}
According to Proposition 3.1 of \cite{Small-GainTheorem}, for the output function $y_{0}$, the inequality
\begin{alignat}{1}
\|y_{0}(t)\|\leq &\gamma_{0}^{u}(\|d_{0}(t)\|_{s})+\gamma_{0}^{y}(\|y_{1}(t)\|_{s})+\beta_{0}(\|x_{0}(0)\|,t)\label{inequality of y0}
\end{alignat}
holds for a pair of class $\mathcal{K}$ functions $(\gamma_{0}^{u},\gamma_{0}^{y})$ and a class $\mathcal{KL}$ function $\beta_{0}$. We have assumed that $g_{1}$ is invertible in a reasonable flight domain, thus, for system $H_{2}$, the virtual control law
\begin{equation}\label{virtual control law 2}
x_{2}=g_{1}^{-1}\left(-f_{1}-\frac{1}{2\delta_{1}^{2}}\eta_{1}-K_{1}\eta_{1}\right)\triangleq x_{2}^{*}
\end{equation}
with $K_{1}>$ and $\delta_{1}>0$ can be also designed based on Theorem \ref{theorem for ISS control law} such that
\begin{alignat}{1}
\|\eta_{1}(t)\|\leq e^{-K_{1}t}\|\eta_{1}(0)\|+\frac{\delta_{1}}{\sqrt{2K_{1}}}\sqrt{1-e^{-2K_{1}t}}\|d_{1}(t)\|_{s}+\frac{\delta_{1}}{\sqrt{2K_{1}}}\sqrt{1-e^{-2K_{1}t}}\|y_{0}(t)\|_{s}\label{inequality of eta1 in the ISS form}
\end{alignat}
and
\begin{alignat}{1}
\|y_{1}(t)\|&\leq\|g_{0}\|\|\eta_{1}^{\#}\|\leq \|g_{0}\|\|\eta_{1}\|\nonumber\\
&\leq \underbrace{\|g_{0}\|\frac{\delta_{1}}{\sqrt{2K_{1}}}\sqrt{1-e^{-2K_{1}t}}\|d_{1}(t)\|_{s}}_{\gamma_{1}^{u}(\|d_{1}(t)\|_{s})}+\underbrace{\|g_{0}\|\frac{\delta_{1}}{\sqrt{2K_{1}}}\sqrt{1-e^{-2K_{1}t}}\|y_{0}(t)\|_{s}}_{\gamma_{1}^{y}(\|y_{0}(t)\|_{s})}+\underbrace{\|g_{0}\|e^{-K_{1}t}\|\eta_{1}(0)\|}_{\beta_{1}(\|\eta_{1}(0)\|,t)}\label{inequality of y1}
\end{alignat}
hold.
Since $\gamma_{1}^{y}\rightarrow0$ as $K_{1}\rightarrow\infty$ or $\delta_{1}\rightarrow0$, Eq. (\ref{requirement of small-gain theorem}) holds for $\gamma_{1y}=\gamma_{0}^{y}$, $\gamma_{2y}=\gamma_{1}^{y}$ and $s_{l}=0$ if proper coefficients $K_{1}$ and $\delta_{1}$ are chosen. In this case, due to Theorem \ref{Small-Gain Theorem}, system $H_{1}$-$H_{2}$ with $(d_{0},d_{1})$ as input, $(y_1,y_2)$ as output and $(x_{0},\eta_{1})$ as state is IOS, and furthermore, it is easy to verify from Eqs. (\ref{inequality of LOS rate in the ISS form in H1}) and (\ref{inequality of eta1 in the ISS form}) that system $H_{1}$-$H_{2}$ is also ISS. Particularly, substituting Eq. (\ref{inequality of y0}) into Eq. (\ref{inequality of y1}) yields\footnote{For any class $\mathcal{K}$ function $\gamma$, any class $\mathcal{K_{\infty}}$ function $\rho$ such that $\rho-Id$ is of class $\mathcal{K_{\infty}}$, and any nonnegative real numbers $a$ and $b$ we have
\begin{equation}\nonumber
\gamma(a+b)\leq \gamma(\rho(a))+\gamma(\rho\circ(\rho-Id)^{-1}(b))
\end{equation}}
\begin{alignat}{1}
\|y_{1}(t)\|_{s}\leq & \gamma_{1}^{u}(\|d_{1}(t)\|_{s})+\gamma_{1}^{y}(\gamma_{0}^{u}(\|d_{0}(t)\|_{s})+\gamma_{0}^{y}(\|y_{1}(t)\|_{s})+\beta_{0}(\|x_{0}(0)\|,0))+\beta_{1}(\|\eta_{1}(0)\|,0)\nonumber\\
\leq & \gamma_{1}^{y}\circ(Id +\rho_{1})\circ\gamma_{0}^{y}(\|y_{1}(t)\|_{s})\nonumber\\
&+\gamma_{1}^{y}\circ(Id +\rho_{1}^{-1})(\gamma_{0}^{u}(\|d_{0}(t)\|_{s})+\beta_{0}(\|x_{0}(0)\|,0))+\gamma_{1}^{u}(\|d_{1}(t)\|_{s})+\beta_{1}(\|\eta_{1}(0)\|,0)\label{inequality of y1 with y0}
\end{alignat}
where $\rho_{1}$ is a class $\mathcal{K}_{\infty}$ function. A fact to be noticed is that if Eq. (\ref{requirement of small-gain theorem}) holds for $\gamma_{1y}=\gamma_{0}^{y}$, $\gamma_{2y}=\gamma_{1}^{y}$ and $s_{l}=0$, the inequality
\begin{eqnarray}\nonumber
\left\{
\begin{aligned}
\gamma_{1}^{y}\circ(Id+\rho_{1})\circ\gamma_{0}^{y}(s)\leq (Id+\rho_{2})^{-1}(s)\\
\gamma_{0}^{y}\circ(Id+\rho_{2})\circ\gamma_{1}^{y}(s)\leq (Id+\rho_{1})^{-1}(s)
\end{aligned}\right.,
\forall s\geq 0
\end{eqnarray}
will hold. Thus,
{\small\begin{alignat}{1}
\|y_{1}(t)\|_{s}
\leq & (Id+\rho_{2})^{-1}(\|y_{1}(t)\|_{s})\nonumber\\
&+\gamma_{1}^{y}\circ(Id +\rho_{1}^{-1})(\gamma_{0}^{u}(\|d_{0}(t)\|_{s})+\beta_{0}(\|x_{0}(0)\|,0))+\gamma_{1}^{u}(\|d_{1}(t)\|_{s})+\beta_{1}(\|\eta_{1}(0)\|,0)\nonumber\\
\leq &(Id+\rho_{2}^{-1})(\gamma_{1}^{y}\circ(Id +\rho_{1}^{-1})(\gamma_{0}^{u}(\|d_{0}(t)\|_{s})+\beta_{0}(\|x_{0}(0)\|,0))+\gamma_{1}^{u}(\|d_{1}(t)\|_{s})+\beta_{1}(\|\eta_{1}(0)\|,0))\label{inequality of y1 wrt d0 & d1}
\end{alignat}}
Substituting the above inequality into Eq. (\ref{inequality of LOS rate in the ISS form in H1}) yields
\begin{alignat}{1}
\|x_{0}(t)\|\leq&\beta_{0}^{x}(\|x_{0}(0)\|,t)+\alpha_{0}^{x}(\|d_{0}(t)\|_{s})\nonumber\\
&+r_{m}\alpha_{0}^{x}((Id+\rho_{2}^{-1})(\gamma_{1}^{y}\circ(Id +\rho_{1}^{-1})(\gamma_{0}^{u}(\|d_{0}(t)\|_{s})+\beta_{0}(\|x_{0}(0)\|,0))\nonumber\\
&+\gamma_{1}^{u}(\|d_{1}(t)\|_{s})+\beta_{1}(\|\eta_{1}(0)\|,0)))\label{inequality of LOS rate in ISS wrt d0 & d1}
\end{alignat}

Next, the small-gain theorem will be used again to propose the final IGC law based on the former design procedures. The change of variables
$$\eta_{2}=x_{2}-x_{2}^{*}$$
brings Eq. (\ref{IGC model}) into the form
\begin{eqnarray}\nonumber
H_{3}:\left\{
\begin{aligned}
\dot{z}&=\begin{bmatrix}
f_{0}+g_{0}x_{1}^{\#}\\
f_{1}-\dot{x}_{1}^{*}
\end{bmatrix}+\begin{bmatrix}
0_{2\times2} & 0_{2\times3}\\
0_{3\times2} & g_{1}
\end{bmatrix}\begin{bmatrix}
0_{2\times1}\\
x_{2}^{*}
\end{bmatrix}+\begin{bmatrix}
0_{2\times1}\\
y_{3}
\end{bmatrix}+\underbrace{\begin{bmatrix}
\frac{d_{0}}{r}\\
d_{1}
\end{bmatrix}}_{d_{3}}\\
y_{2}&=-\dot{x}_{2}^{*}
\end{aligned}\right.
\end{eqnarray}
and
\begin{eqnarray}\nonumber
H_{4}:\left\{
\begin{aligned}
\dot{\eta}_{2}&=f_{2}+g_{2}u+d_{2}+y_{2}\\
y_{3}&=g_{1}\eta_{2}
\end{aligned}\right.
\end{eqnarray}
where $z=[x_{0}^{T},\eta_{1}^{T}]^{T}$. As a result of the former analysis, with $x_{2}^{*}$, system $H_{3}$ is ISS with respect to $y_{3}$ and $d_{3}$, and particularly, from Eq. (\ref{inequality of LOS rate in ISS wrt d0 & d1}), we have
\begin{alignat}{1}
\|x_{0}(t)\|\leq&\beta_{0}^{x}(\|x_{0}(0)\|,t)+\alpha_{0}^{x}(\|d_{0}(t)\|_{s})\nonumber\\
&+r_{m}\alpha_{0}^{x}((Id+\rho_{2}^{-1})(\gamma_{1}^{y}\circ(Id +\rho_{1}^{-1})(\gamma_{0}^{u}(\|d_{0}(t)\|_{s})+\beta_{0}(\|x_{0}(0)\|,0))\nonumber\\
&+\gamma_{1}^{u}(\|d_{1}(t)\|_{s}+\|y_{3}(t)\|_{s})+\beta_{1}(\|\eta_{1}(0)\|,0)))\label{inequality of LOS rate in ISS wrt d0 & d1 & y3}
\end{alignat}
For output function $y_{2}$, the inequality
\begin{equation}
\|y_{2}(t)\|\leq \gamma_2^y(\|y_{3}(t)\|_s)+\gamma_2^u(\|d_{3}(t)\|_s)+\beta_2(\|z(0)\|,t)
\end{equation}
holds for a pair of $\mathcal{K}$ functions $(\gamma_2^y,\gamma_2^u)$ and a class $\mathcal{KL}$ function $\beta_2$.
For system $H_{4}$, we can also design a controller based on Theorem \ref{theorem for ISS control law} as follows
\begin{equation}\label{IGC law}
u=g_{2}^{-1}\left(-f_{2}-\frac{1}{2\delta_{2}^{2}}\eta_{2}-K_{2}\eta_{2}\right)
\end{equation}
and the following inequalities hold
\begin{equation}
\|\eta_{2}(t)\|\leq e^{-K_{2}t}\|\eta_{2}(0)\|+\frac{\delta_{2}}{\sqrt{2K_{2}}}\sqrt{1-e^{-2K_{2}t}}\|d_{2}(t)\|_{s}+\frac{\delta_{2}}{\sqrt{2K_{2}}}\sqrt{1-e^{-2K_{2}t}}\|y_{2}(t)\|_{s}
\end{equation}
\begin{alignat}{1}
\|y_{3}(t)\|&\leq\|g_{3}\|\|\eta_{2}\|\leq \|g_{3}\|\|\eta_{2}\|\nonumber\\
&\leq \underbrace{\|g_{3}\|\frac{\delta_{2}}{\sqrt{2K_{2}}}\sqrt{1-e^{-2K_{2}t}}\|d_{2}(t)\|_{s}}_{\gamma_{3}^{u}(\|d_{2}(t)\|_{s})}+\underbrace{\|g_{3}\|\frac{\delta_{2}}{\sqrt{2K_{2}}}\sqrt{1-e^{-2K_{2}t}}\|y_{2}(t)\|_{s}}_{\gamma_{3}^{y}(\|y_{2}(t)\|_{s})}+\underbrace{\|g_{3}\|e^{-K_{2}t}\|\eta_{2}(0)\|}_{\beta_{3}(\|\eta_{2}(0)\|,t)}\label{inequality of y3}
\end{alignat}
Due to the small-gain theorem and the form of $\gamma_{3}^{y}$, we know that if small enough $\delta_{2}$ or big enough $K_{2}$ is used, Eq. (\ref{requirement of small-gain theorem}) will hold for $\gamma_{1y}=\gamma_{2}^{y}$, $\gamma_{2y}=\gamma_{3}^{y}$ and $s_{l}=0$, that is, $H_{3}$ and $H_{4}$ is IOS with respect to $d_{2}$ and $d_{3}$. Moreover, similarly to the procedure from Eq. (\ref{inequality of y1 with y0}) to Eq. (\ref{inequality of y1 wrt d0 & d1}), the inequality
\begin{equation}
\|y_{3}(t)\|_{s}\leq (Id+\rho_{2}^{-1})(\gamma_{3}^{y}\circ(Id +\rho_{1}^{-1})(\gamma_{2}^{u}(\|d_{3}(t)\|_{s})+\beta_{2}(\|z(0)\|,0))+\gamma_{3}^{u}(\|d_{2}(t)\|_{s})+\beta_{3}(\|\eta_{2}(0)\|,0))\label{inequality of y3 wrt d2 & d3}
\end{equation}
can be obtained. Substituting the above inequality into Eq. (\ref{inequality of LOS rate in ISS wrt d0 & d1 & y3}) yields
\begin{alignat}{1}
\|x_{0}(t)\|\leq&\beta_{0}^{x}(\|x_{0}(0)\|,t)+\alpha_{0}^{x}(\|d_{0}(t)\|_{s})\nonumber\\
&+r_{m}\alpha_{0}^{x}((Id+\rho_{2}^{-1})(\gamma_{1}^{y}\circ(Id +\rho_{1}^{-1})(\gamma_{0}^{u}(\|d_{0}(t)\|_{s})+\beta_{0}(\|x_{0}(0)\|,0))\nonumber\\
&+\gamma_{1}^{u}(\|d_{1}(t)\|_{s}+(Id+\rho_{2}^{-1})(\gamma_{3}^{y}\circ(Id +\rho_{1}^{-1})(\gamma_{2}^{u}(\|d_{3}(t)\|_{s})+\beta_{2}(\|z(0)\|,0))\nonumber\\
&+\gamma_{3}^{u}(\|d_{2}(t)\|_{s})+\beta_{3}(\|\eta_{2}(0)\|,0)))+\beta_{1}(\|\eta_{1}(0)\|,0)))\label{inequality of LOS rate in ISS wrt d0 & d1 & d2}
\end{alignat}
Since $\gamma_{1}^{u},\gamma_{1}^{y}\rightarrow0$ as $K_{1}\rightarrow\infty$ or $\delta_{1}\rightarrow0$ and $\gamma_{3}^{u},\gamma_{3}^{y}\rightarrow0$ as $K_{2}\rightarrow\infty$ or $\delta_{2}\rightarrow0$, it can be seen that the right-hand
side of (\ref{inequality of LOS rate in ISS wrt d0 & d1 & d2}) approaches
\begin{equation}
\beta_{0}^{x}(\|x_{0}(0)\|,t)+\alpha_{0}^{x}(\|d_{0}(t)\|_{s})+r_{m}\alpha_{0}^{x}\circ(Id+\rho_{2}^{-1})(\beta_{1}(\|\eta_{1}(0)\|,0))
\end{equation}
as $K_{1},K_{2}\rightarrow\infty$ or $\delta_{1},\delta_{2}\rightarrow0$ for bounded $d_{0}$, $d_{1}$ and $d_{2}$, which shows that for sufficiently small $\delta_{1},\delta_{2}$ or sufficiently big $K_{1},K_{2}$ the influence of $d_{1}$ and $d_{2}$ on $x_{0}$ will be close to zero. Besides that, due to the form of $\alpha_{0}^{x}$, $d_{0}$ and $\beta_{1}(\|\eta_{1}(0)\|,0)$ can be also suppressed by adjusting $K_{0}$ and $\delta_{0}$.

Thus, the main results can be summarized as the following theorem.
\begin{theorem}
Consider the guidance and control system (\ref{IGC model}). Assume that $g_{0}(t)$ and $g_{1}(\vartheta,x_{1})$ are invertible in a reasonable flight domain. For bounded $d_{i}(t)~(i=0,1,2)$, the IGC law
\begin{eqnarray}
\left\{
\begin{aligned}
x_{1}^{\#*}&=g_{0}^{-1}\left(2\frac{V_{r}}{r}-\frac{1}{2\delta^{2}_{0}}-K_{0}\right)x_{0}\\
\eta_{1}&=x_{1}-[0,(x_{1}^{\#*})^{T}]^{T}\\
x_{2}^{*}&=g_{1}^{-1}\left(-f_{1}-\frac{1}{2\delta_{1}^{2}}\eta_{1}-K_{1}\eta_{1}\right)\\
\eta_{2}&=x_{2}-x_{2}^{*}\\
u&=g_{2}^{-1}\left(-f_{2}-\frac{1}{2\delta_{2}^{2}}\eta_{2}-K_{2}\eta_{2}\right)
\end{aligned}\right.
\end{eqnarray}
with positive coefficients $K_{i}$ and $\delta_{i}$ for $i=0,1,2$ can make the variables $x_{0}$, $\eta_{1}$ and $\eta_{2}$ be ISS with respect to $d_{i}~(i=0,1,2)$, and the LOS rate $x_{0}$ can converge into a neighborhood of zero
whose size can be reduced by adjusting the coefficients $K_{i}$ and $\delta_{i}$.
\end{theorem}

\begin{remark}
\cite{Adaptive_block_dynamic_surface_control} introduced a set of first-order filters at each step of the traditional
block backstepping approach to avoid the problem of ``explosion of complexity'', which made the IGC law be complex in structure. Comparing with that method, the structure of our approach is more concise.
\end{remark}

\section{Conclusions}\label{Conclusions}
This paper proposes a three-dimensional integrated guidance and control (IGC) approach by using small-gain theorem. The couplings between the guidance system and control system and those between different channels of the pursuer dynamics are fully and explicitly considered in the design procedure, and our IGC law can guarantee stability of the overall system including the guidance and control loop without the assumption that the angle between LOS and pursuer velocity is almost invariable. Theoretical analysis also shows that the IGC approach can make the line-of-sight (LOS) rate converge into a small neighborhood of zero, and besides, the law is more concise in structure when compared with the existing results.


\begin{thebibliography}{0}
\bibitem{Adaptive_block_dynamic_surface_control}
Hou, M. Z, Liang, X. L, and Duan, G. R. ``Adaptive block dynamic surface control for
integrated missile guidance and autopilot,'' \textit{Chinese Journal of Aeronautics}, Vol. 26, No. 3, 2013, pp. 741-750.
\bibitem{inegrated_guidance_control_first}
Williams, D. E., Richman, J., and Friedland, B., ``Design of an integrated strapdown guidance and control system for a tactical missile,'' AIAA paper 1983-2169, 1983.
\bibitem{SOSM_guidance_control_2009}
Shtessel, Y. B., Shkolnikov, I. A., and Levant, A., ``Guidance and control of missile interceptor using second-order sliding modes,'' \textit{IEEE Transactions on Aerospace and Electronic Systems}, Vol. 45, No.1, 2009, pp. 110-124.
\bibitem{finite_time_inegrated_guidance_control_JGCD}
Shtessel, Y. B., and Tournes, C., ``Integrated higher-order sliding mode guidance and autopilot for dual-control missiles,'' \textit{Journal of Guidance, Control, and Dynamics}, Vol. 45, No. 2, 2009, pp. 110-124.
\bibitem{SSOSM-based_guidance_law}
Shtessel, Y. B., Shkolnikov, I. A., and Levant, A., ``Smooth second-order sliding modes: Missile guidance application,'' \textit{Automatica}, Vol. 43, No. 8, 2007, pp. 1470-1476.
\bibitem{WXH_2013}
Wang, X. H. and Wang, J. Z., ``Partial integrated missile guidance and control with finite time convergence,'' \textit{Journal of Guidance Control and Dynamics}, Vol. 36, No.5, 2013, pp. 1399-1409.
\bibitem{sliding-mode_gl_2006}
Shima, T., Idan, M., and Golan, O. M., ``Sliding-mode control for integrated missile autopilot guidance,'' \textit{Journal of Guidance Control and Dynamics}, Vol. 29, No. 2, 2006, pp. 250-260.
\bibitem{sliding-mode_gl_2007}
Idan, M., Shima, T., and Golan, O. M., ``Integrated sliding mode autopilot-guidance for dual-control missiles,'' \textit{Journal of Guidance Control and Dynamics}, Vol. 30, No. 4, 2007, pp. 1081-1089.
\bibitem{Yan_IGC_backstepping}
Yan, H., Wang, X. H., Yu, B. F., and Ji, H. B.``Adaptive integrated guidance and control based on backstepping and input-to-state stability,'' \textit{Asian Journal of Control}, 2013, doi: 10.1002/asjc.682.
\bibitem{small_gain_theorem_for_IGC}
Yan, H. and Ji, H. B. ``Integrated guidance and control for dual-control missiles based on small-gain theorem,'' \textit{Automatica}, Vol. 48, No. 10, 2012, pp. 2686-2692.
\bibitem{Small-GainTheorem}
Jiang, Z. P., Teel, A. R., and Praly, L., ``Small-gain theorem for ISS systems and applications.'' \textit{Mathematics of Control, Signals, and Systems}, Vol. 7, No. 2, 1994, pp. 95-120.
\bibitem{Riccati_for_inegrated_guidance_control_by_Menon}
Menon, P. K., and Ohlmeyer, E. J., ``Integrated design of agile missile guidance and control systems'', \textit{Proceedings of the 7th Mediterranean Conference on Control and Automation}, Haifa, 1999, pp. 1469-1494.
\bibitem{Riccati_for_inegrated_guidance_control_by_palumbo}
Palumbo, N. F., and Jackson, T. D., ``Integrated missile guidance and control:
A state dependent Riccati differential equation approach,'' \textit{Proceedings of International Conference on Control Applications}, Hawai'i, 1999, pp. 243-248.
\bibitem{theta-D_method_for_inegrated_guidance_control}
Xin, M., Balakrishnan, S. N., and Ohlmeyer, E. J., ``Integrated guidance and control of missiles
with $\theta$-D method,'' \textit{IEEE Transactions on Control Systems Technology}, Vol. 14, No.6, 2006, pp. 981-992.
\bibitem{Smooth_stabilization_implies_coprime_factorization}
Sontag, E. D., ``Smooth stabilization implies coprime factorization,'' \textit{IEEE Transactions on Automatic Control}, Vol. 34, No. 4, 1989, pp. 435-443.
\bibitem{book_by_Li}
Li, X. G., Fang, Q.,\textit{Winged missile flight dynamics}, Xi¡¯an: Northwestern Polytechnical University Press, 2004 (in Chinese).
\bibitem{Yan_HGOGL}
Yan, H. and Ji, H. B.,``Guidance laws based on input-to-state stability and high-gain observers,'' \textit{IEEE Transactions on Aerospace and Electronic Systems}, Vol. 48, No. 3, 2012, pp. 2518-2529.
\end{thebibliography}
\end{document}